\documentclass[sigconf]{acmart}
\AtBeginDocument{%
  }

\copyrightyear{2026}
\acmYear{2026}
\setcopyright{cc}
\setcctype{by}
\acmConference[SIGIR '26]{Proceedings of the 49th International ACM SIGIR Conference on Research and Development in Information Retrieval}{July 20--24, 2026}{Melbourne, VIC, Australia}
\acmBooktitle{Proceedings of the 49th International ACM SIGIR Conference on Research and Development in Information Retrieval (SIGIR '26), July 20--24, 2026, Melbourne, VIC, Australia}
\acmDOI{10.1145/3805712.3809894}
\acmISBN{979-8-4007-2599-9/2026/07}
  
\usepackage[table]{xcolor}
\usepackage{booktabs}
\usepackage{cleveref}
\usepackage{pifont}
\newcommand{\cmark}{\ding{51}}%
\newcommand{\xmark}{\ding{55}}%

\usepackage[title]{appendix}

\makeatletter
\def\NAT@def@citea{\def\@citea{\NAT@separator}}
\makeatother

\definecolor{lightred}{RGB}{246, 216, 219}
\definecolor{lightgreen}{RGB}{200, 231, 204}
\definecolor{lightblue}{RGB}{198, 215, 253}



\title{Semantic Recall for Vector Search}

\author{Leonardo Kuffo}
\affiliation{%
  \institution{CWI}
  \city{Amsterdam}
  \country{The Netherlands}}
\email{lxkr@cwi.nl}

\author{Ioanna Tsakalidou}
\affiliation{%
    \institution{EPFL}
    \city{Lausanne}
    \country{Switzerland}
}
\email{ioanna.tsakalidou@epfl.ch}

\author{Roberta De Viti}
\affiliation{%
    \institution{MPI-SWS}
    \city{Saarbrücken}
    \country{Germany}
}
\email{rdeviti@mpi-sws.org}

\author{Albert Angel}
\affiliation{%
    \institution{Google}
    \city{Zurich}
    \country{Switzerland}
}
\email{albertangel@google.com}

\author{Jiří Iša}
\affiliation{%
    \institution{Google}
    \city{Zurich}
    \country{Switzerland}
}
\email{jisa@google.com}

\author{Rastislav Lenhardt}
\affiliation{%
    \institution{Google}
    \city{Zurich}
    \country{Switzerland}
}
\email{lenhardt@google.com}

\crefname{section}{\S\!}{\S\S}
\crefname{appendix}{\S\!}{\S\S}
\crefname{figure}{Figure}{Figures}
\crefname{table}{Table}{Tables}

\begin{abstract}

We introduce \textit{Semantic Recall}, a novel metric to assess the quality of 
approximate nearest neighbor search algorithms by considering only semantically relevant objects that are theoretically retrievable via exact nearest neighbor search. Unlike traditional recall, semantic recall does not penalize algorithms for failing to retrieve objects that are semantically irrelevant to the query, even if those objects are among their nearest neighbors. We demonstrate that semantic recall is particularly useful for assessing retrieval quality on queries that have few relevant results among their nearest neighbors---a scenario we uncover to be common within embedding datasets. Additionally, we introduce \textit{Tolerant Recall}, a proxy metric that approximates semantic recall when semantically relevant objects cannot be identified. We empirically show that our metrics are more effective indicators of retrieval quality, and that optimizing search algorithms for these metrics can lead to improved cost-quality tradeoffs.

\end{abstract}

\begin{document}

\maketitle

\begin{table*}[t!]
\renewcommand{\tabcolsep}{5.0pt}
\centering
\caption{Comparison of quality metrics for ANNS.}
\vspace*{-4mm}
\resizebox{0.8\linewidth}{!}{%
\begin{tabular}{lccccccc}
\hline
\textbf{Criteria} 
& \textbf{\cellcolor{lightred}{\begin{tabular}[c]{@{}c@{}}Traditional\\ Recall\end{tabular}}} 
& \textbf{\begin{tabular}[c]{@{}c@{}} Using curated \\ neighbors\end{tabular}} 
& \textbf{\cellcolor{lightblue}{\begin{tabular}[c]{@{}c@{}}Semantic\\ Recall\end{tabular}}} 
& \textbf{\cellcolor{lightgreen}{\begin{tabular}[c]{@{}c@{}}Tolerant \\ Recall\end{tabular}}} 
& \textbf{\begin{tabular}[c]{@{}c@{}}Recall\\ $@k$-$\epsilon$\end{tabular}} 
& \textbf{\begin{tabular}[c]{@{}c@{}}Relative \\ Distance Error\end{tabular}} 
& \textbf{R-Precision} \\ 
\hline

\begin{tabular}[c]{@{}l@{}}Measures semantic relevance\end{tabular} 
& \xmark & \cmark & \cmark & \begin{tabular}[c]{@{}c@{}} Partial \end{tabular} & \xmark & \xmark & \cmark \\ 
\hline

\begin{tabular}[c]{@{}l@{}}Robust to model errors \end{tabular} 
& \cmark & \xmark & \cmark & \cmark & \cmark & \cmark & \xmark \\ 
\hline

\begin{tabular}[c]{@{}l@{}} Hyperparameter-free \end{tabular} 
& \cmark & \cmark & \cmark & \xmark & \xmark & \cmark & \cmark \\ 
\hline

\begin{tabular}[c]{@{}l@{}} No manual curation required \end{tabular} 
& \cmark & \xmark & \begin{tabular}[c]{@{}c@{}} Partial \end{tabular} & \cmark & \cmark & \cmark & Partial \\ 
\hline

\begin{tabular}[c]{@{}l@{}} Robust to dataset updates \end{tabular} 
& \xmark & \xmark & \xmark & \cmark & \cmark & \cmark & \xmark \\ 
\hline

Interpretability of score 
& Easy & Easy & Easy & Fair & Fair & Hard & Fair \\ 
\hline

\end{tabular}
\label{tab:comparison}
}
\end{table*}

\section{Introduction}

In recent years, information retrieval (IR) using vector embeddings has received unprecedented attention from both academia and industry. This surge is fueled by AI embedding models that effectively capture concepts from objects (e.g., text, images, videos) into high-dimensional vectors~\cite{team2023gemini, geminiembedding, mxbai}. In high-dimensional vector spaces, the proximity of two vectors in terms of their semantic concepts is captured by distance metrics like Euclidean distance or inner product~\cite{gao2021simcse, reimers2019sentence}. Consequently, modern IR systems switched to \textit{semantic search}: user queries are transformed into vector embeddings, which are used to identify their nearest objects in a database using nearest neighbor search (NNS) algorithms~\cite{vectordbsurvey, scann, faisslibrary}. 

Exact NNS is computationally expensive; to accelerate it, IR systems relax the constraints of search \textit{exactness} and provide only \textit{approximate} results. While these approximations are sufficient for most IR tasks, they raise a question: \textit{How to measure search quality?}~\cite{integrityjunkiness}. Ideally, developers would use end-to-end quality metrics during the development of approximate NNS (ANNS) algorithms (e.g., click-through rates, satisfaction surveys). However, this is generally impractical~\cite{cabitza2023toward, ding2022posthoc}; thus, academia and industry have standardized the use of \textit{recall} for assessing retrieval quality of ANNS algorithms~\cite{scann, scannalloydb, quake, darth, vibebenchmark, micronn, pdx}. Given a dataset and a set of query embeddings, \textit{recall} is defined as the proportion of retrieved nearest neighbors appearing in the queries' \textit{ground truth}. A ground truth is a set of ``true" nearest neighbors identified by ranking all dataset embeddings via brute-force (exact) NNS. 

\begin{figure}[t]
\centering
\includegraphics[width=0.85\columnwidth]{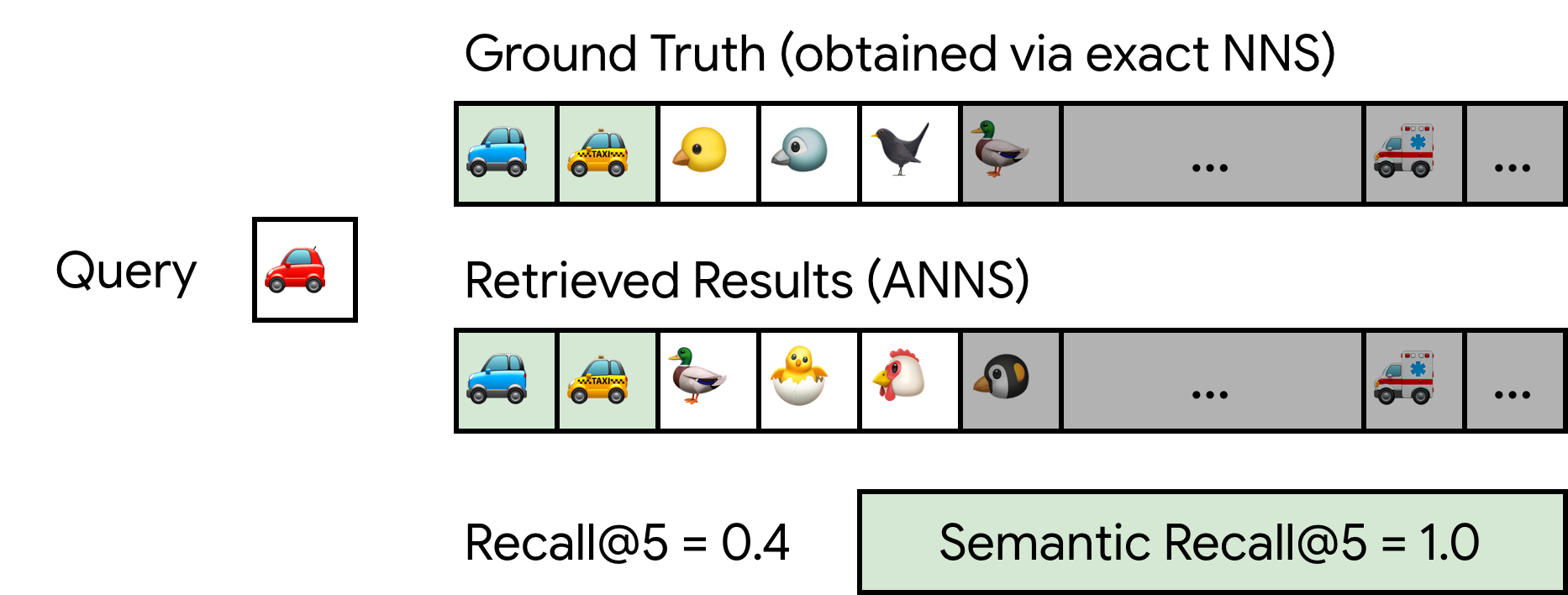}
\vspace*{-2.0mm}
\caption{Semantic recall only considers the semantically relevant results in the ground truth that are theoretically retrievable via exact nearest neighbor search (in green).}
\label{fig_opening}
\vspace*{-3.0mm}
\end{figure}


Despite its widespread use, recall fails to accurately capture the semantic quality of retrieved results~\cite{darth, integrityjunkiness, robustness, vibebenchmark, chen2025reveal}. This limitation stems from the reliance on ground truth defined via exact NNS, which reflects mathematical proximity rather than semantic relevance. As a result, low recall in ANNS can be misleading: failing to retrieve a mathematically close neighbor is penalized even when that neighbor is semantically irrelevant. Essentially, traditional recall penalizes the omission of “mathematical noise”: irrelevant points that happen to lie closer to the query in the embedding space. This issue is particularly prevalent when queries have few relevant results, a scenario we find to be common in embedding datasets (\cref{subsection_analysis_semantic}). 
Consequently, traditional recall is misaligned with the goal of embedding-based retrieval: identifying semantically similar objects.

Occasionally, datasets include a curated ground truth where queries are manually matched to relevant objects. However, ANNS algorithms cannot retrieve \textit{semantically} better neighbors than those identified by an exact NNS, as ANNS is constrained by the concepts captured in the embedding space~\cite{limitationsembeddings}. For instance, in Figure~\ref{fig_opening}, the ambulance emoji is an item relevant to the query. 
However, the embedding model failed to place this object close to the query in the vector space. Thus, an exact NNS misses it, and, by extension, any ANNS algorithm. Still, a curated ground truth may recognize the item as a relevant neighbor to the query. As a result, the use of curated ground truths may inadvertently penalize ANNS algorithms for limitations of the embedding model~\cite{chen2025reveal}.

To address these issues, we introduce two novel metrics to evaluate retrieval quality of ANNS algorithms: semantic recall and tolerant recall. We define \textbf{semantic recall} as the fraction of \textit{semantic neighbors} retrieved by an ANNS algorithm that are \textit{theoretically retrievable} via exact NNS (Figure~\ref{fig_opening}). Semantic neighbors are the subset of the ground truth that exhibit a semantic relationship with the query, as determined by an external \textit{judge}. Furthermore, we introduce \textbf{tolerant recall}, a metric that approximates semantic recall by allowing the replacement of a true nearest neighbor with another retrieved result if their scores are \textit{close} (i.e., within a given tolerance).


In the remainder of this work, we make the following contributions: (i) The definition of semantic recall, our novel proxy for search quality of ANNS algorithms (\cref{sec:srecall}); (ii) A  methodology for identifying the semantic neighbors of a set of queries (\cref{subsec_sn}); (iii) The introduction of tolerant recall, a metric for scenarios where semantic neighbors cannot be obtained (\cref{sec:trecall}); (iv) An empirical comparison of traditional recall, semantic recall, and tolerant recall (\cref{sec:exp}); (v) An empirical evaluation showing potential for improving the performance and reliability of ANNS systems by tuning search algorithms with our metrics (\cref{subsection_tuning}).

\section{Semantic Recall}
\label{sec:srecall}

We define semantic neighbors as the subset of brute-force ground truth neighbors that are \textit{semantically relevant} to the query. This semantic relevance is determined by an external \textit{judge} (\cref{subsec_sn}). Given a set of semantic neighbors (SN) taken from the top-$k$ true nearest neighbors of a query, and a set of neighbors (R) retrieved by an ANNS algorithm, we define semantic recall (srecall) as the portion of R in SN: \textit{srecall} = $\frac{| R \cap SN |}{|SN|}$. A high srecall indicates that the ANNS algorithm is effective at retrieving objects that are not just mathematically close, but also semantically meaningful. 

Semantic recall addresses three critical limitations of traditional recall: (i) it does not penalize algorithms for missing mathematically closer but semantically irrelevant objects;
(ii) it circumvents the limitations of the embedding model by focusing only on relevant objects that are present in the ground truth obtained via exact NNS, and (iii) it eliminates the need to configure threshold hyperparameters. This stands in contrast with metrics like $recall@k\text{-}\epsilon$ (used in ANN-benchmarks~\cite{annbenchmarks}), $robustness@k\text{-}\gamma$~\cite{robustness}, and Relative Distance Error (RDE)~\cite{darth, rabitqext}, among others~\cite{integrityjunkiness}. These metrics require thresholds that lack clear configuration heuristics and suffer from poor interpretability and portability across datasets. Moreover, RDE still penalizes ANNS algorithms for missing irrelevant neighbors. Finally, we believe that semantic recall is a useful metric for both end-user applications and developers. As a more accurate indicator of retrieval quality, it is a better proxy to measure whether an algorithmic improvement has any real impact on the end-user experience.



\subsection{Finding Semantic Neighbors}
\label{subsec_sn}

To identify semantic neighbors, we employ a two-step process. First, we compute the ground truth of the query embeddings via exact NNS. Then, we classify each object in the ground truth as either semantically relevant or irrelevant to the query. This classification can be performed by human judges or by Large Language Models (LLMs), which excel at classification tasks~\cite{devlin2019bert, raffel2020exploring}. The judging process with LLMs entails engineering a prompt tailored to the retrieval workload and the dataset (e.g., for Q\&A corpora, a relevant document should answer the query text). Notably, this approach is significantly more efficient than dataset curation: instead of judging millions of (query, document) pairs to create a ``gold standard`` dataset, we only validate a reduced subset per query (i.e., the top-$k$ results in the ground truth).

\section{Tolerant Recall}
\label{sec:trecall}


In some scenarios (e.g., when one has only embeddings but not the underlying data), it may not be possible to identify semantic neighbors. To this end, we propose tolerant recall, a metric that allows a retrieved result to replace a true nearest neighbor when measuring search quality, if the distance of the query from this result is \textit{close} to the distance of the query from the true neighbor. 

Let us assume that we search for 20 ground-truth documents ($G_{20}$) and we retrieve 20 documents ($R_{20}$). Tolerant recall (trecall) identifies the largest subset $T \subset R_{20}$ s.t. every object $t_i \in T$ corresponds to a unique object $g_i \in G_{20}$, where either $g_i = t_i$, i.e., the document is in the ground truth, or, assuming inner product similarity, $t_i^{score} \geq g_i^{score} \cdot (1-x\%$), i.e., the scores of these documents are within a $x\%$ tolerance threshold. Then: $trecall@20 = \frac{|T|}{|G_{20}|}$. 

As shown in \cref{subsection_traditional_semantic_tolerant}, tolerant recall closely tracks semantic recall, as it is robust to small perturbations in the distance metric that alter the ranking of irrelevant neighbors, such as those introduced by quantization~\cite{rabitqext}. Furthermore, tolerant recall is more robust for monitoring frequently-updated datasets, whereas traditional recall can rapidly degrade if the ground truth is not updated alongside each database update. Additionally, for situations like a query having two high-scoring semantic results and the rest of the results being irrelevant (low-scoring results), tolerant recall captures the case of losing one of the semantic results. In contrast, $recall@k\text{-}\epsilon$~\cite{annbenchmarks} may not register this as a loss, since it counts a result as correct if its score falls within a ($1 +\epsilon$) threshold of the $k$-th neighbor in the ground truth. 

Table~\ref{tab:comparison} compares various recall metrics, highlighting the benefits of semantic and tolerant recall for evaluating ANNS algorithms.

\section{Evaluation}
\label{sec:exp}

We evaluate our metrics on the MSMARCO corpus, which contains 8.83M text documents. We encode all documents into 3072-dimensional embeddings using the Gemini Embedding model~\cite{geminiembedding}. Our query set consists of 250 general-knowledge questions, encoded with the same model. For each query, we compute the top-100 ground-truth neighbors via brute-force search using inner product similarity. Since the embeddings are normalized, inner product scores lie in $[-1,1]$, with higher values indicating greater similarity. To identify semantic neighbors, we use Gemini~2.5~\cite{comanici2025gemini} as a judge to label each ground-truth neighbor as either \emph{Relevant} or \emph{Not Relevant} with respect to the query.

\subsection{Analysis of Semantic Neighbors}
\label{subsection_analysis_semantic}

\begin{figure}
\includegraphics[width=0.80\columnwidth]{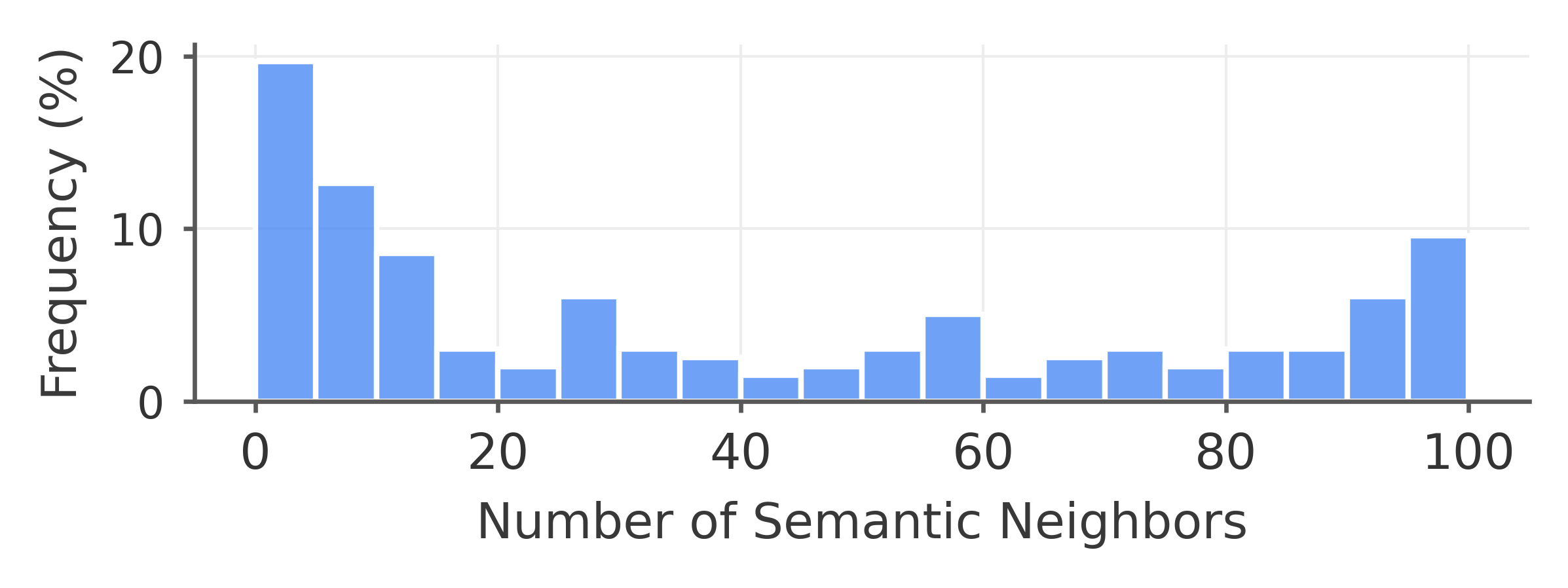}
\centering
\vspace*{-5.0mm}
\caption{Distribution of semantically relevant neighbors per query in the MSMARCO dataset (250 queries).}
\label{fig_distribution_semantic}
\end{figure}

Figure~\ref{fig_distribution_semantic} shows the distribution of semantic neighbors per query as identified by Gemini. The distribution is bimodal: 40\% of the queries have few relevant documents (0-15), including 3 queries with no relevant results, while another cluster appears near the upper end of the range (90-100 relevant documents). Further inspection reveals that queries in this latter group have more than 100 relevant documents in the corpus; the apparent upper bound is due to only using the top-100 neighbors in our experiments. 

\begin{table}[t!]
\renewcommand{\tabcolsep}{3.0pt}
\centering
\caption{Inner product scores for semantic neighbors (SN) and non-semantic neighbors (non-SN) across queries.}
\vspace*{-4mm}
\resizebox{\linewidth}{!}{%
\begin{tabular}{lcccc}
\toprule
& \textbf{All Queries} & \textbf{SN < 20} & \textbf{20 $\leq$ SN < 80} & \textbf{SN $\geq$ 80} \\
\midrule
N. Queries & 247 & 102 & 91 & 54 \\ \hline
Avg. score (SN)  & 0.77 ($\pm$ 0.02) & \textbf{0.77 ($\pm$0.03)} & 0.76 ($\pm$0.02) & 0.77 ($\pm$0.01) \\
Avg. score (non-SN) & 0.73 ($\pm$0.03) & \textbf{0.71 ($\pm$0.02)} & 0.74 ($\pm$0.02) & 0.76 ($\pm$0.02) \\ \hline

Avg. score deltas (SN)  & 0.006 ($\pm$0.006) & \textbf{0.014 ($\pm$0.012)} & 0.002 ($\pm$0.003) & 0.001 ($\pm$0.001) \\
Avg. score deltas (non-SN) & 0.002 ($\pm$0.003) & \textbf{0.001 ($\pm$0.002)} & 0.001 ($\pm$0.002) & 0.005 ($\pm$0.006) \\
\bottomrule
\end{tabular}%
\label{table_dp}
\vspace*{-2mm}
}
\end{table}

Table~\ref{table_dp} compares the inner product scores of semantic neighbors (SN) and non-semantic neighbors (non-SN). SNs achieve higher average similarity scores across all query groups, with the largest gap appearing for queries with fewer than 20 SNs. Notably, SNs exhibit considerably larger and more variable score differences (deltas) between consecutive neighbors. In contrast, non-SNs exhibit consistently small score deltas, indicating that most irrelevant documents lie at nearly the same distance from the query embedding. This concentration increases sensitivity to ranking fluctuations among irrelevant ground-truth neighbors, increasing the likelihood that ANNS algorithms may omit such neighbors, leading to penalties under traditional recall~\cite{vibebenchmark}. 
These findings underscore the necessity for semantic recall: strictly penalizing algorithms for retrieving one irrelevant document over another is counterproductive.



\subsubsection{Cross validation} 
We cross-validated a subset of relevance judgments using Claude Haiku 4.5 as an independent judge outside the Gemini family. Specifically, we sampled 100 queries and their top-100 ground truth documents. The agreement rate between judges was 91.1\% (Cohen's $\kappa$ = 0.82). Looking into each category, agreement was 89.5\% for "Relevant" and 93.3\% for "Not Relevant". Only 2 queries fell below 70\% agreement rate. 

When the judges disagreed, Gemini more frequently labeled documents as "Relevant", while Claude adhered more strictly to the prompt criteria. Subsequent human inspection suggested that Gemini was more lenient, reasoning about ambiguities in query phrasing and document intent. For example, for the query \textit{"when is the warm weather in Thailand"} and the document \textit{"Thailand can best be described as tropical and humid for the majority of the country during most of the year."}, Claude labeled the document as not relevant due to the lack of specific temporal information, whereas Gemini labeled it as relevant.
These findings highlight the importance of designing prompts that fit the use case of the application. For reference, our prompt is available in our code notebook~\cite{codereprod}. 


\subsection{Traditional, Semantic, and Tolerant Recall}
\label{subsection_traditional_semantic_tolerant}

\begin{table}[t!]
\renewcommand{\tabcolsep}{3.0pt}
\centering
\caption{Summary of traditional, semantic, and tolerant recall on MSMARCO.}
\vspace*{-4mm}
\label{tab:recall_comparison}
\resizebox{\linewidth}{!}{%
\begin{tabular}{l cc cc cc}
\toprule
& \multicolumn{2}{c}{\cellcolor{lightred}Traditional Recall} & \multicolumn{2}{c}{\cellcolor{lightblue}Semantic Recall} & \multicolumn{2}{c}{\cellcolor{lightgreen}Tolerant Recall} \\
\cmidrule(lr){2-3} \cmidrule(lr){4-5} \cmidrule(lr){6-7}
 & Avg. & Std. & Avg. & Std. & Avg. & Std. \\
\midrule
All queries & 0.863 & 0.177 & \textbf{0.932} & 0.144 & \textbf{0.920} & 0.125 \\
Queries with < 20 SN & 0.762 & 0.204 & \textbf{0.903} & 0.183 & \textbf{0.859} & 0.152 \\
Queries with 20 to 80 SN & 0.903 & 0.133 & \textbf{0.937} & 0.124 & \textbf{0.941} & 0.096 \\
Queries with $\geq$ 80 SN & 0.976 & 0.061 & \textbf{0.978} & 0.062 & \textbf{0.991} & 0.027 \\
\bottomrule
\end{tabular}
\label{table_summary}
}
\vspace*{-2.0mm}
\end{table}

\begin{figure}[t!]
\includegraphics[width=1.0\columnwidth]{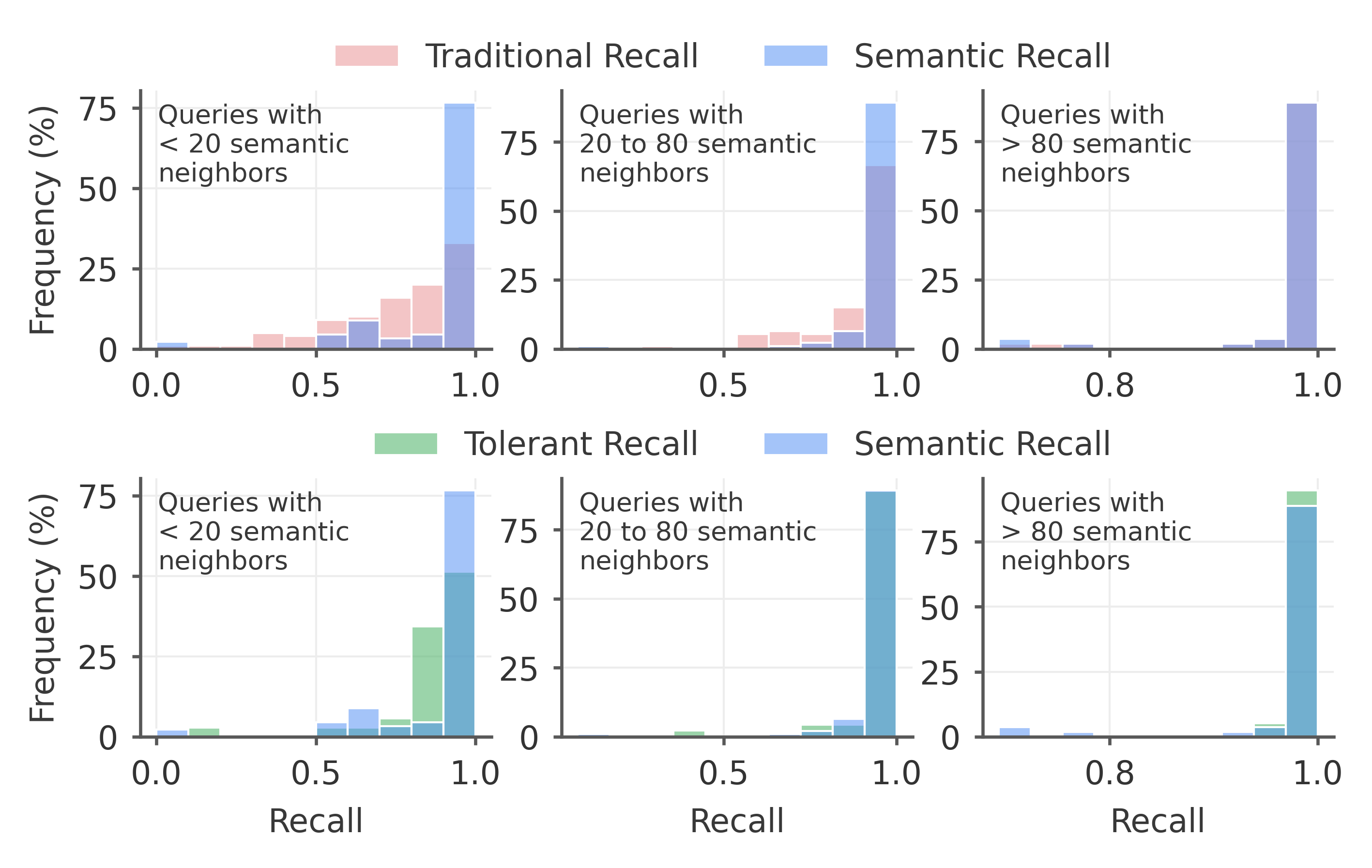}
\centering
\vspace*{-9.0mm}
\caption{Per-query distributions of traditional (red), semantic (blue), and tolerant (green) recall on the MSMARCO dataset, grouped by semantic neighbor count (left to right).}
\label{fig_recalls}
\vspace*{-2.0mm}
\end{figure}

To evaluate traditional, semantic, and tolerant recall, we ran top-100 searches using ScANN~\cite{scann, scannalloydb} (a partition-based ANNS index), with 8-bit quantization. Table~\ref{table_summary} summarizes the results, and Figure~\ref{fig_recalls} (top) shows the per-query distribution of traditional and semantic recall over 250 queries. Queries with few SNs are the most penalized by traditional recall, as near-equidistant irrelevant neighbors lead to missed ground-truth matches. In contrast, semantic recall better reflects the search quality perceived by end users (i.e., queriers), yielding a score of \colorbox{lightblue}{0.903} compared to \colorbox{lightred}{0.762} for queries with few SN. This pattern holds in top-20 search, where traditional recall scores \colorbox{lightred}{0.869} and semantic recall \colorbox{lightblue}{0.901}. 
Tolerant recall with a 1\% tolerance threshold closely approximates semantic recall (Figure~\ref{fig_recalls}, bottom). 
Unlike semantic recall, tolerant recall can be directly integrated into existing ANNS pipelines and remains usable in dynamic datasets. 

\begin{figure}[h]
  \includegraphics[width=1.0\columnwidth]{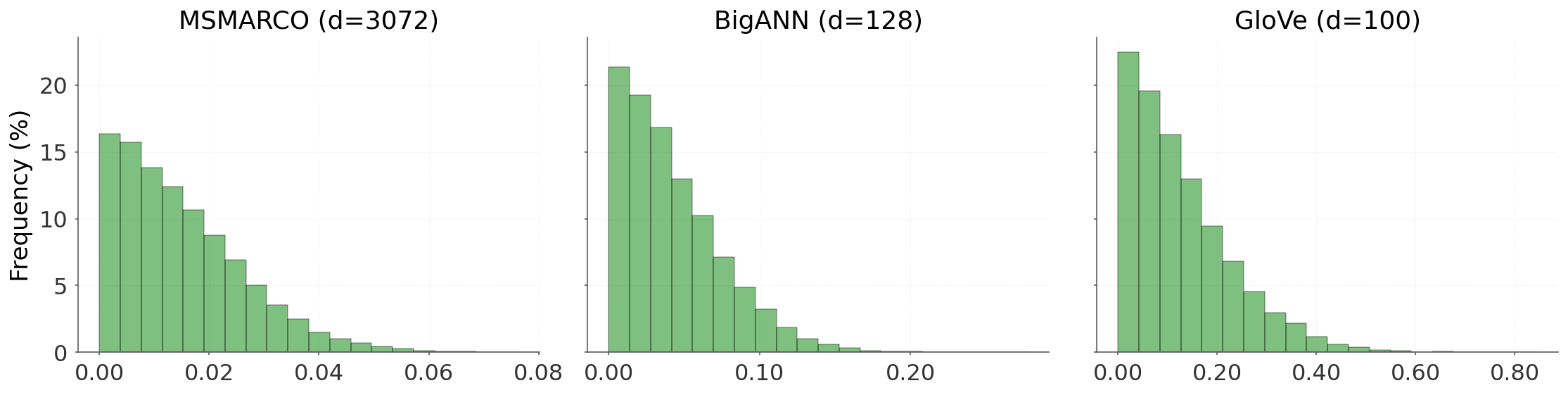}
  \centering
  \vspace*{-7.0mm}
  \caption{Distribution of the error \% between scores computed with vectors at full precision and 8-bit quantization. }
  \label{fig:quanterrrors}
\vspace*{-2mm}
\end{figure}

Notably, the closeness of non-SNs also negatively affects traditional recall when evaluating quantization techniques. Figure~\ref{fig:quanterrrors} shows the relative errors between inner product scores computed with full precision and with 8-bit scalar quantization across MSMARCO, GloVe~\cite{pennington2014glove}, and BigANN~\cite{simhadri2022results}. Even these small errors can reorder irrelevant neighbors, leading to penalties under traditional recall.
Tolerant recall mitigates this effect by allowing replacements mostly within irrelevant results, and semantic recall avoids it by considering only relevant results, which are typically top-rankers in the ground truth.

\subsubsection{Comparison to R-Precision} 


Given the identified SNs, retrieval quality could be evaluated using R-precision. R-precision assumes that for a query with $R$ relevant objects, an ideal retrieval system retrieves all $R$ within the top-$R$ results~\cite{rprecision}. However, R-precision inadvertently penalizes the retrieval performance of ANNS algorithms for shortcomings in the embedding model. In fact, it assumes that the ground-truth ranking is aligned with semantic relevance, ignoring that irrelevant items may appear among top-ranked neighbors~\cite{agarwal2026strengths}. Although ground-truth rank and semantic relevance are negatively correlated for queries with fewer than 20 SNs (rank-biserial correlation of $-0.72$, p-value$\approx0$), this correlation is not perfect. For instance, when retrieving $k=5$ neighbors with $R=3$ SNs, if the third SN is ranked fourth in the ground truth, R-precision counts this as an error. However, in this case, even an exact NNS would not have retrieved that document within the top-$R$.
In contrast, semantic recall overcomes this limitation by allowing relevant documents to appear anywhere within the top-$k$ results.


\subsection{Tuning for Semantic Recall}
\label{subsection_tuning}
ANNS systems are commonly tuned to balance recall and search cost by selecting Pareto-optimal hyperparameters~\cite{faisslibrary,annbenchmarks,vibebenchmark}. We show that using semantic and tolerant recall as tuning objectives yields a strictly better cost–quality tradeoff than tuning for traditional recall.
We use Google Vizier~\cite{vizier} to tune the hyperparameters of ScaNN, minimizing search cost at a fixed target recall. We measure search cost as ``bytes read'', which closely aligns with the number of inner product computations~\cite{scannautotune}. Finally, we assess \textit{performance} by measuring the CPU cost of query processing. 

First, we tune ScaNN for a target traditional recall of 98\% on MSMARCO, obtaining a configuration of Pareto-optimal hyperparameters that achieves 98\% traditional recall, \colorbox{lightgreen}{98.69\%} tolerant recall, and \colorbox{lightblue}{98.93\%} semantic recall. Then, we re-tune with tolerant recall as the objective, targeting \colorbox{lightgreen}{98.69\%}: this yields a configuration with the same tolerant recall at 5\% lower cost. Similarly, tuning for semantic recall and targeting \colorbox{lightblue}{98.93\%} reduces cost by 14\% while preserving semantic recall.
These gains arise because non-SNs are harder to retrieve: in partition-based indexes such as ScaNN, they require exploring additional partitions, increasing search effort.

\begin{figure}[t!]
\includegraphics[width=0.60\columnwidth]{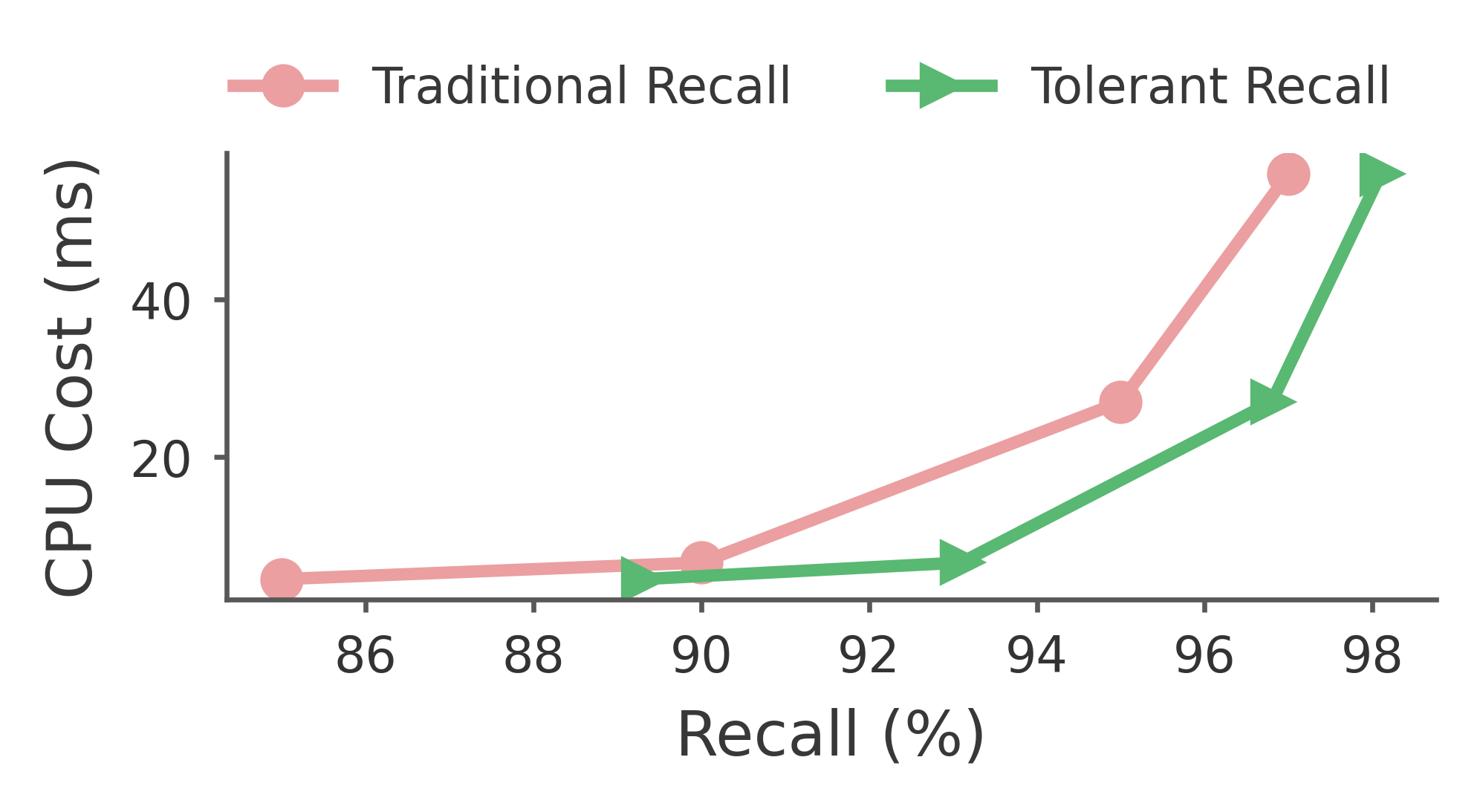}
\centering
\vspace*{-5.0mm}
\caption{Recall vs cost: Cost rises sharply as recall increases.}
\vspace*{-2.0mm}
\label{fig:cost_curve}
\end{figure}

Finally, we observe a significant opportunity to reduce costs by changing the target metric, e.g., from 95\% traditional recall to 95\% tolerant recall. The latter metric being a better representative of the real losses and always higher, enables this. In particular, tuning for 95\% tolerant recall instead of 95\% traditional recall reduces cost by \textasciitilde 25\% on BigANN (1B vectors, 128 dimensions), \textasciitilde 5\% on GloVe (1M vectors, 100 dimensions), and \textasciitilde 35\% on MSMARCO. Notably, as shown in Figure~\ref{fig:cost_curve}, the cost grows sharply at higher recalls. As a result, whenever we manage to move slightly left on the recall-cost curve, we substantially reduce compute costs.

\section{Generalizability}\label{sec:general}
\begin{figure}[t!]
\includegraphics[width=0.65\columnwidth]{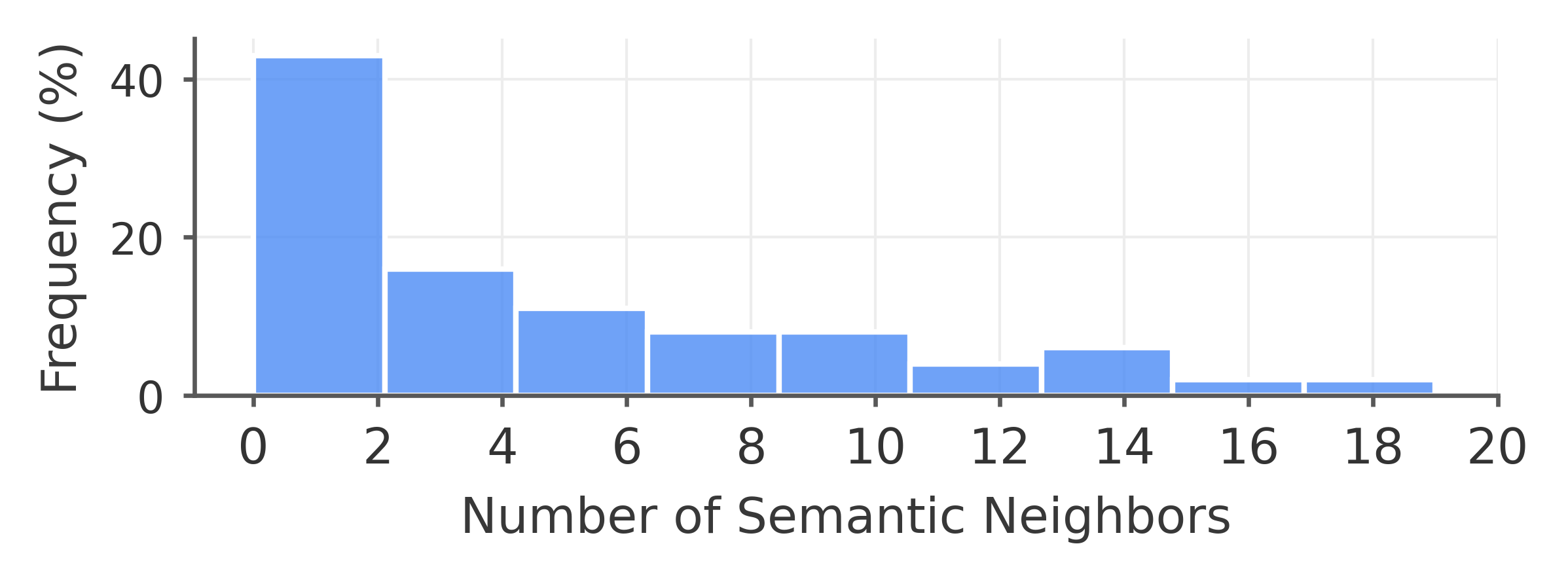}
\centering
\vspace*{-5.0mm}
\caption{Distribution of the number of semantic neighbors per query in the MIRACL dataset.}
\vspace*{-4.0mm}
\label{fig_miracl_distribution_semantic}
\end{figure}

\begin{figure}[t!]
\includegraphics[width=1.0\columnwidth]{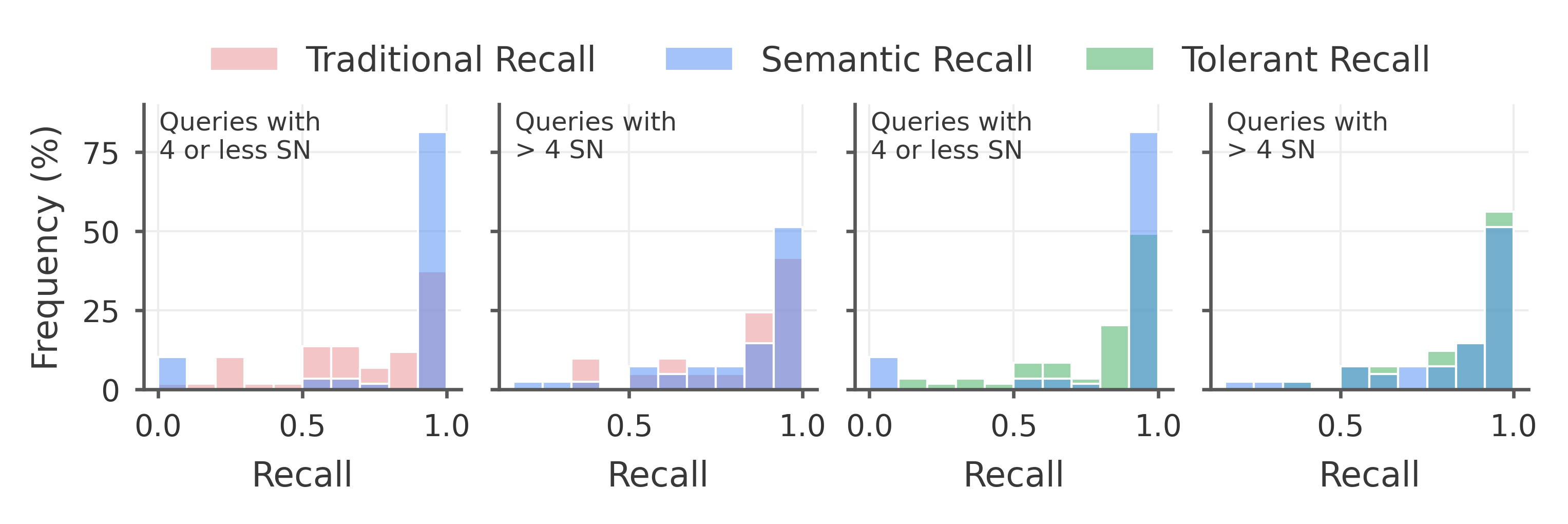}
\centering
\vspace*{-9.0mm}
\caption{Per-query distribution of traditional, semantic, and tolerant recall on the MIRACL dataset.}
\label{fig_miracl_recalls}
\vspace*{-2.0mm}
\end{figure}

To demonstrate the generalizability of our metrics, we replicate our evaluation on a subset of the MIRACL dataset~\cite{zhang2023miracl} consisting of 542,166 Thai-language documents. Documents and queries are embedded into 1024-dimensional vectors using Cohere Embed V3~\cite{cohere2023introducing}. The query set comprises 100 questions in Thai. For each query, we compute the top-20 ground truth neighbors via brute-force inner product search. We identify semantic neighbors (SNs) using Gemini 2.5~\cite{comanici2025gemini} as a judge. Figure~\ref{fig_miracl_distribution_semantic} shows the distribution of SNs per query, which exhibits a power-law behavior similar to MSMARCO: 60\% of the queries have at most 4 SNs. Similar to the MSMARCO dataset, SNs exhibit larger score deltas than non-SNs (0.05$\pm$0.03 in SN vs. 0.005$\pm$0.008 in non-SN). We then evaluate traditional, semantic, and tolerant recall with a 2\% tolerance threshold using a FAISS IVF index~\cite{faisslibrary} (3000 partitions, 8-bit scalar quantization), probing 0.5\% of clusters. The average traditional, semantic, and tolerant recall are \colorbox{lightred}{0.75}, \colorbox{lightblue}{0.85}, and \colorbox{lightgreen}{0.83}, respectively. Figure~\ref{fig_miracl_recalls} shows per-query recall distributions. As in MSMARCO, queries with few SNs exhibit low traditional recall, while semantic recall confirms that relevant results are retrieved. Tolerant recall serves as a middle ground, providing a better approximation of quality without requiring semantic judgments. 

\subsection{Choosing a tolerance threshold} 
\begin{figure}[t!]
\includegraphics[width=1.0\columnwidth]{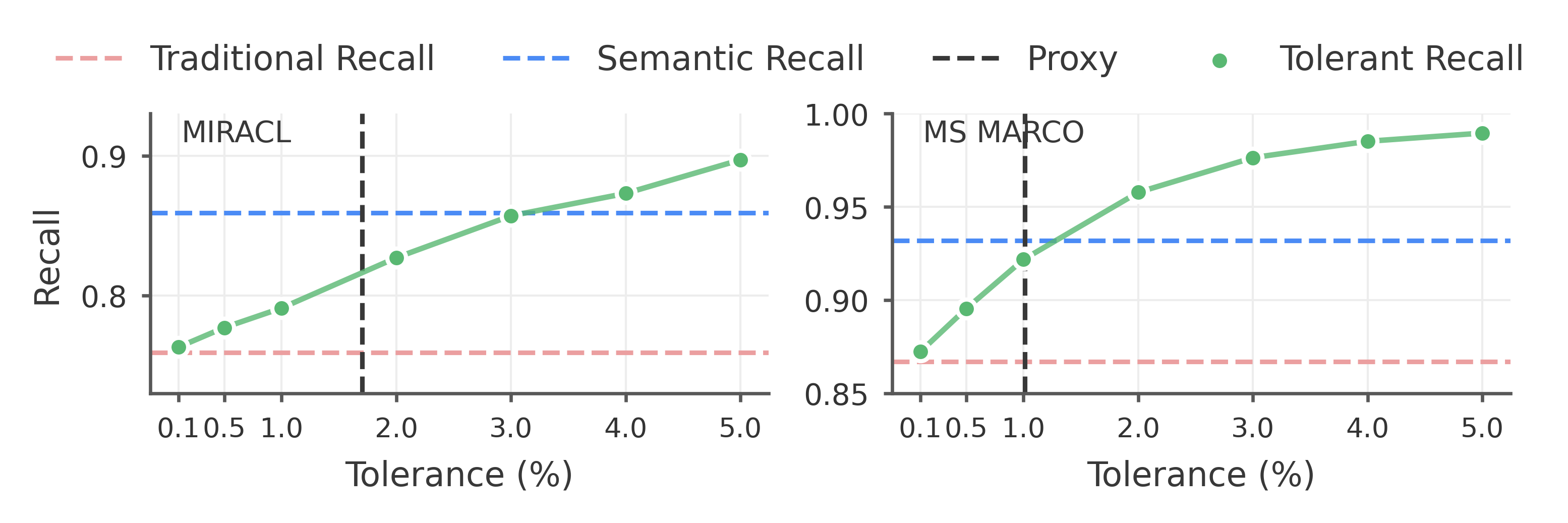}
\centering
\vspace*{-9.0mm}
\caption{Effect of different thresholds on tolerant recall. Our proxy for setting a threshold closely follows semantic recall.}
\vspace*{-4.0mm}
\label{fig_tolerance}
\end{figure}

Tolerant recall allows replacing irrelevant documents with retrieved documents with \textit{similar} scores. Accordingly, the tolerance threshold should be sufficiently broad to enable such substitutions. We approximate this threshold using score differences among lower-ranked neighbors. In particular, we compute the relative difference between the $\frac{2}{3}k$-th and $k$-th neighbors from a top-$k$ retrieval, normalized by the maximum ground-truth score. The tolerance threshold proxy is thus defined as $ \frac{score_{2/3 k} - score_{k}}{  max\_{score}} * 100$ averaged across queries. Figure~\ref{fig_tolerance} shows that varying the threshold leads to different tolerant recall values. In both datasets, our proxy for the tolerance threshold closely matches the semantic recall metric. While this provides a useful starting point, there is potential for improvement, e.g., by estimating thresholds per query rather than per dataset. A conservative lower bound for the tolerance threshold can be determined from the relative error of scores resulting from the 8-bit compression of vectors, as 8 bits are sufficient to achieve almost perfect traditional recall in ANNS~\cite{rabitqext}. Finally, when SNs are available, the tolerance threshold can alternatively be obtained from the intersection of the tolerance threshold curve with semantic recall.

\section{Discussion}

Our results suggest that ANNS evaluation can relax strict mathematical precision in favor of improved search efficiency without compromising retrieval quality more than previously thought. 
In fact, semantic recall enables developers to distinguish between performance losses that affect relevant results and those that reshuffle irrelevant ones, supporting more meaningful cost---quality tradeoffs during algorithm design (e.g., when tuning quantization schemes or search parameters~\cite{darth}).
We further hypothesize that the long-tailed distributions of traditional recall reported in prior work on ANNS~\cite{vibebenchmark,darth} stem from queries with few or no relevant neighbors. In such cases, hyperparameter tuning selects ``costly" configurations that incur into additional effort to recover near-equidistant \textit{irrelevant} neighbors, effectively optimizing for ``noise".
Thus, optimizing for traditional recall means optimizing for mathematical exactness at unnecessary extra retrieval cost. 

\paragraph{Limitations.}
Semantic recall has practical limitations. Unlike traditional recall, it requires access to the underlying raw objects (text, images, ...) for judging. Also, the judging relies on the LLM's ``knowledge": domain-specific datasets may still require human validation. However, judging only the top-$k$ ground truth is more feasible than annotating entire datasets and more useful than pooled annotations. Furthermore, our approach adopts binary relevance judgments, which collapse uncertainty into a hard decision and delegate the confidence threshold to the judge. While richer graded judgments could capture uncertainty, we favor a binary definition for simplicity and reproducibility.
Errors in judgment are unavoidable, and correlations between the judge and the evaluated system may occur in some scenarios. 
Finally, semantic recall is undefined when no relevant results exist (e.g., highly restrictive queries). Identifying these queries through the judging process is useful and worth inspection, as it may indicate regression in the embedding model. However, we argue that such queries should be excluded from retrieval-quality evaluation, as no meaningful answers are present. Conversely, when all results are relevant, semantic recall generalizes to recall, whereas tolerant recall enables replacements with objects that are likely also relevant due to their score proximity. 

\section{Conclusions}

We introduced \textit{semantic} and \textit{tolerant} recall as alternatives to traditional recall that better capture retrieval quality in semantic search. Our results show that traditional recall penalizes ANNS algorithms for failing to retrieve near-equidistant but semantically irrelevant results, which are difficult to distinguish and costly to recover.
By accounting for this structure, semantic and tolerant recall provide more faithful evaluation metrics and enable more meaningful cost–quality tradeoffs for algorithm design and tuning. Lastly, we validated the effectiveness of our metrics across different embedding models, ANNS algorithms, and data scales, demonstrating their applicability to vector search systems. We hope this work encourages the IR and Vector Search communities to incorporate evaluation metrics that better align with the objective of semantic search: retrieving relevant answers rather than the mathematically closest noise. 
As future work, we propose incorporating our metrics into early-termination mechanisms~\cite{darth,quake} and comparing our metrics across different ANNS algorithms (e.g., HNSW~\cite{hnsw}). We provide a code notebook to reproduce our experiments~\cite{codereprod}.


\balance
\bibliographystyle{ACM-Reference-Format}
\bibliography{main}

\end{document}